\documentclass[preprint,showpacs,preprintnumbers,amsmath,amssymb]{revtex4}

\usepackage[pdftex]{graphicx}% Include figure files
\usepackage{graphicx}% Include figure files
\usepackage{dcolumn}% Align table columns on decimal point
\usepackage{bm}% bold math

\begin{document}

\title{Dephasing of quantum dot exciton polaritons in electrically tunable nanocavities}% Force line breaks with \\

\author{A. Laucht, N. Hauke, J. M. Villas-B\^{o}as, F. Hofbauer, M. Kaniber, G. B\"{o}hm}
\author{J. J. Finley}%
\email{finley@wsi.tum.de}
\affiliation{%
Walter Schottky Institut, Technische Universit\"at M\"unchen, Am Coulombwall 3, D-85748 Garching, Germany
}%

\date{\today}% It is always \today, today,

\begin{abstract}
We experimentally and theoretically investigate dephasing of zero dimensional microcavity polaritons in electrically tunable single dot photonic crystal nanocavities. Such devices allow us to alter the dot-cavity detuning \emph{in-situ} and to directly probe the influence on the emission spectrum of varying the incoherent excitation level and the lattice temperature.  By comparing our results with theory we obtain the polariton dephasing rate and clarify its dependence on optical excitation power and lattice temperature. For low excitation levels we observe a linear temperature dependence, indicative of phonon mediated polariton dephasing. At higher excitation levels, excitation induced dephasing is observed due to coupling to the solid-state environment. The results provide new information on coherence properties of quantum dot microcavity polaritons.
\end{abstract}

%Use showkeys class option if keyword
\maketitle
%INTRODUCTION
Two distinct regimes of light-matter interaction exist in cavity-QED; weak coupling in which excited emitters decay irreversibly and strong coupling (SC), where a coherent exchange of energy occurs between the emitter and a single mode of the quantized electromagnetic field\cite{Andreani99,Khitrova06}. For single semiconductor quantum dots (QDs) in solid state microcavities strong coupling was first observed a few years ago\cite{Reithmaier04,Yoshie04}, the key signature being an anticrossing between an exciton ($\omega_{x}$) and the cavity mode ($\omega_{c}$) as they are tuned into resonance ($\delta=\omega_{x}-\omega_{c}=0$)\cite{Andreani99,Khitrova06}.
%The SC regime is of strong interst since it directly provides an optical non-linearity at the single photon level, a property that may be used to construct single photon transistors\cite{Bermel06,Chang07} or even mediate coherent coupling between spatially separated quantum systems\cite{Bouwmeester00}.\\
Surprisingly, recent theoretical work has shown that an anticrossing in the emission spectrum is \emph{not} an unambiguous signature of SC\cite{Laussy08a,Laussy08b}, rather its appearance depends on the balance between the incoherent pumping rates of the excitonic ($P_{x}$) and photonic ($P_{c}$) sub systems\footnote{The dot is pumped by the continuous laser excitation into the excited state continuum and the cavity field by other dots that non-resonantly emit photons into the resonator\cite{Kaniber08b}.}.
The appearance of an anticrossing then depends on whether the steady quantum state of the system is more photon-like or exciton-like\cite{Laussy08a,Laussy08b} and also on the importance of pure dephasing. Clearly, probing the nature of the light-matter coupling in such systems calls for a method to measure the emission spectrum as a function of $\delta$ under well defined experimental conditions of optical pumping and lattice temperature. Comparison of the results with theory can then be used to extract quantitative information on the mechanisms responsible for dephasing. To date $\delta$ has been tuned by varying the lattice temperature\cite{Reithmaier04,Yoshie04,Peter05,Press07,Englund07,Faraon08} or by condensing inert gases into the cavity at low temperatures\cite{Mosor05,Hennessy07,Winger08}. Varying the lattice temperature simultaneously affects the exciton dephasing rate\cite{Borri05} \emph{and} $\delta$, greatly complicating the comparison of experiment with theory. Similarly, inert gas deposition can only be used for $T\leq25$~K\cite{Mosor05} making detailed temperature dependent studies of dephasing impossible.\\
In this letter we measure the emission spectrum ($S(\omega)$) of an electrically tunable single dot nanocavity operating in the strong coupling regime under controlled experimental conditions. Such devices allow us to vary $\delta$ \emph{in-situ} using the quantum confined Stark effect\cite{Finley04,Laucht09} and to probe the systematic evolution of $S(\omega)$ with the laser power and lattice temperature.  By comparing our results with a modified version (including pure dephasing) of a recently published theory\cite{Laussy08a,Laussy08b} we obtain the polariton dephasing rate and its dependence on lattice temperature and the level of incoherent optical excitation. For a fixed excitation power we observe a linear temperature variation of the dephasing rate indicative of a phonon mediated process. In addition, at a fixed lattice temperature we observe excitation induced dephasing due to coupling of the 0D polariton states to the solid-state environment. The results obtained provide information on the processes that mediate dephasing of 0D-microcavity polaritons.\\

%\begin{figure}[b!]
%\includegraphics[width=86.3mm]{figure1}
%\caption{\label{figure1} (color online) Typical photoluminescence spectra recorded as function of the applied bias voltage showing different single dot transitions (labelled 1,2,3,4) that shift due to the DC Stark effect and the cavity mode. \textbf{(inset)} Schematic cross-sectional representation of the device and layer sequence of the active region, showing the underetched region of the PC membrane. The polarity of the static electric field ($F$) is indicated.}
%\end{figure}

%TEXT FOR FIGURE 1
%As depicted schematically in the inset of Fig. 1 
The samples investigated were GaAs $p$-$i$-$n$ photodiodes with membrane type PC nanocavities patterned into the active layers\cite{Hofbauer07, Laucht09}. Low density InGaAs self-assembled QDs ($\leq20$ $\mu$m$^{-2}$) were incorporated into the middle of the intrinsic region, allowing electric fields to be applied by varying the voltage across the junction ($V_{app}$). %Electrical contacts were established to the $n-$ and $p$-doped contact layers using optical lithography and $250\times400$ $\mu$m rectangular photodiode mesas were formed using wet chemical etching. An array of 5$\times$5 hexagonal PCs were defined into the top-contact window using electron beam lithography and reactive ion etching, before the membranes were underetched using $HF$ acid\cite{Hofbauer07}. A modified $L3$ defect was included \cite{Akahane03} at the center of each photonic crystal to create high-Q nanocavities. 
The fabricated $L3$ nanocavities\cite{Akahane03} support six strongly localized modes within the 2D-photonic bandgap of which the fundamental cavity mode exhibits Q-factors ranging from $8000$ to $12000$ in our structures, high enough to reach the SC regime\cite{Reithmaier04,Yoshie04}.\\
Photoluminescence spectra were recorded from one of the nanocavities using confocal microscopy as a function of $V_{app}$. The excitation laser was tuned to a higher energy cavity mode in order to select only dots spatially coupled to the fundamental cavity mode\cite{Nomura06,Kaniber09}. Using the quantum confined Stark effect\cite{Laucht09}, one of the QD transitions can be shifted through the cavity mode ($\hbar\omega_{c}=1207.1\pm0.1$ meV) at $V_{app}\sim-0.17$~V, a voltage where the QD emission intensity remains constant since the onset of carrier tunnelling escape from the dots occurs at larger reverse biases\cite{Laucht09}.

%TEXT FOR FIGURE 2
A detailed high resolution voltage sweep of this situation is presented in the left panel of Fig. 1 for $T=18$ K and $P_{exc}=25$ W/cm$^{2}$. A clear anticrossing is observed between the exciton and cavity mode as they are electrically tuned into resonance.  The open blue circles correspond to the experimental data whilst the solid black lines are fits using the theoretical model described below. The right panel of Fig. 1 shows a magnified image of three of these spectra to demonstrate the excellent agreement between the measured and calculated spectral functions.

\begin{figure}[t!]
\includegraphics[width=86.3mm]{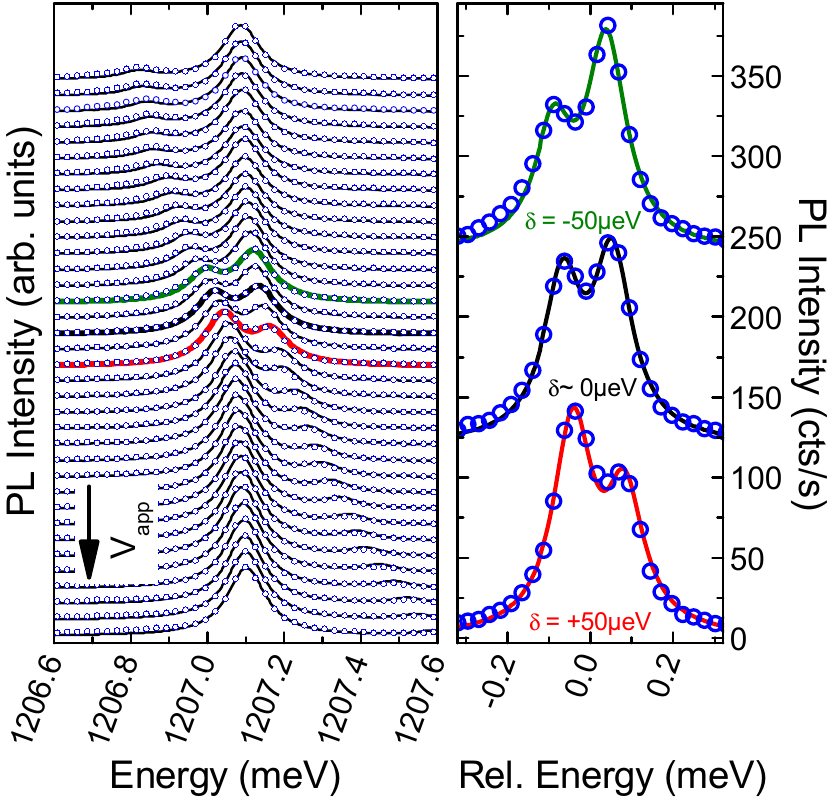}
\caption{\label{figure2} (color online) \textbf{(left panel)} Waterfall plot of exciton-cavity anticrossing as $V_{app}$ is changed from $-0.05$V to $-0.40$V, corresponding to detunings ranging from $\delta=+450\mu$eV to $\delta=-260\mu$eV. The open circles correspond to experimental data whilst the solid lines are fits to the theory. Selected curves close to resonance ($\delta=0\mu$eV) are presented in the right panel showing the excellent agreement between the observed and calculated spectral functions.}
\end{figure}

To extract quantitative information on dephasing of the 0D polaritons we simulated the spectral function of the dot-cavity system using an extended form of the model published in Refs. \cite{Carmichael89,Laussy08a,Laussy08b}. We use the following Hamiltonian to account for the exciton-cavity interaction
\begin{equation}
    H=\frac{\hbar\omega_x}{2}\sigma_z+\hbar\omega_c a^\dagger a +\hbar g (a^\dagger\sigma_- + \sigma_+ a),
\end{equation}
where $\sigma_+$, $\sigma_-$ and $\sigma_z$ are the pseudospin operators for the two level system consisting of the QD ground state $|0\rangle$ and a single exciton $|X\rangle$ state, $\omega_x$ is the exciton frequency, $a^\dagger$ and $a$ are the creation and destruction operators of photons in the cavity mode with frequency $\omega_c$, and $g$ is the dipole coupling between cavity mode and exciton. The incoherent loss and gain (pumping) of the dot-cavity system is then included in the master equation of the Lindblad form $d\rho/dt=-i/\hbar[H,\rho]+\mathcal{L}(\rho)$,
with the super-operator
\begin{eqnarray*}
&&\hspace{-0.3cm}\mathcal{L}(\rho)=\frac{\Gamma_x}{2}(2\sigma_-\rho \sigma_+ - \sigma_+ \sigma_-\rho - \rho \sigma_+ \sigma_-) \\&&\hspace{-0.3cm}+ \frac{P_x}{2}(2\sigma_+\rho \sigma_- - \sigma_- \sigma_+\rho - \rho \sigma_- \sigma_+)+ \frac{\gamma_x^\phi}{2}(2\sigma_z\rho \sigma_z - \rho )\\&&\hspace{-0.3cm}+\frac{\Gamma_c}{2}(2a\rho a^\dagger - a^\dagger a\rho - \rho a^\dagger a )+\frac{P_c}{2}(2a^\dagger \rho a - a a^\dagger\rho - \rho a a^\dagger).
\end{eqnarray*}
Here, $\Gamma_x$ is the exciton decay rate, $P_x$ is the rate in which excitons are created by the continuous wave pump laser tuned to the higher order cavity  mode, $\Gamma_c$ is the cavity loss rate and $P_c$ is the incoherent pumping of the cavity mode\footnote{Pumping of the cavity from non-resonant QDs was observed and investigated from different groups as \cite{Press07, Hennessy07, Kaniber08b}. The mechanism responsible is still a topic of ongoing experiments and does not form part of this publication.}. $\gamma_x^\phi$ is the pure dephasing rate of the exciton, needed to describe effects originating from high excitation powers or high temperatures as discussed below.

Assuming that most of the light collected in our setup escapes the system through the leakage of the cavity and using the Wiener-Khintchine theorem, the spectral function of the collected light is given by $S(\omega)\propto\lim_{t\rightarrow\infty}\mathrm{Re}\int_0^\infty d\tau e^{-(\Gamma_r-i\omega)\tau}\langle a^\dagger(t)a(t+\tau)\rangle$\cite{Scully97}, where the term $\hbar\Gamma_r=30$ $\mu$eV (half-width) was added to take into account the finite spectral resolution of our Triax-550 imaging spectrometer with 1200 lines/mm grating \cite{Eberly77}.
%Here, we have assumed that the system achieves a steady state for long times. To compute the two-time correlation function for the photon operator we derive the Liouvillian equations for the average operators, which can be separated in two sets of differential equations and make use of the quantum regression theorem as in Ref. \cite{Carmichael89}. 
The spectral function with a normalization factor $N$ is obtained analytically using the quantum regression theorem, as in Ref. \cite{Carmichael89}, and is given by
\begin{equation}
    S(\omega)=N\mathrm{Re}\left[\frac{A_+}{i\lambda_+ -i\omega+\Gamma_r}+\frac{A_-}{i\lambda_- -i\omega+\Gamma_r}\right],
\end{equation}
where $A_\pm=[(\omega\pm i(\lambda_x-\lambda_c)/2)\langle a^\dagger a\rangle_{ss}\pm g \langle a^\dagger \sigma_-\rangle_{ss}]/2\Omega$, with emission frequency $ \lambda_\pm=(\lambda_c+\lambda_x)/2\pm \Omega$
and Rabi splitting%
\begin{eqnarray}
\Omega&=&\sqrt{g^2-(\lambda_x-\lambda_c)^2/4},
\end{eqnarray}
where $\lambda_x$ and $\lambda_c$ are defined as
\begin{eqnarray*} 
\lambda_x&=&i\omega_x+(2\gamma_x^\phi+\Gamma_x+P_x)/2,\\ \lambda_c&=&i\omega_c+(\Gamma_c-P_c)/2 .
\end{eqnarray*}
In these equations $\langle a^\dagger a\rangle_{ss}$ and $\langle a^\dagger \sigma_-\rangle_{ss}$ are the steady solution ($t\rightarrow\infty$) of the expected operators.  These solutions are also obtained analytically using a method similar to that presented in Ref. \cite{Laussy08b}.

% \begin{eqnarray}
%     \lambda_\pm=\frac{\omega_c+\omega_0}{2}-i\frac{\gamma_c+\gamma_d}{2}\pm \frac{\sqrt{4g^2+(\Delta-i(\gamma_d-\gamma_c))^2}}{2}
% \end{eqnarray}
% with $\gamma_c = (\Gamma_c-P_c)/2, \gamma_d = \gamma_d^\phi+(\Gamma_d+P_d)/2$, and $\Delta=\omega_d-\omega_c$.

Using this spectral function the set of curves of Fig. 1 was fitted simultaneously using an Levenberg-Marquardt algorithm. All parameters were optimized globally except $N$ which was optimized for each curve individually to compensate for drift of the experimental setup. The best fit for the data presented in Fig. 1 was obtained for $\hbar g=59$ $\mu$eV (which was then held constant for all fittings), $\hbar\Gamma_{x}=0.2$ $\mu$eV, $\hbar P_{x}=0.5$ $\mu$eV, $\hbar\Gamma_{c}=68.0$ $\mu$eV, $\hbar P_{c}=4.5$ $\mu$eV, and $\hbar\gamma_{x}^\phi=19.9$ $\mu$eV. These numbers already highlight that pure dephasing plays a significant role even for the lowest excitation power densities ($\sim1$ $W/cm^{2}$)) and temperatures ($T=18$ $K$) investigated in our experiments.

% while the normalization constant was optimized for each curve individually to be able to compensate for a potential change in the experimental setup.

%TEXT FOR FIGURE 3
Being able to reproduce the measured spectra from theory with very good agreement, and to extract the different parameters from the fits, allows us to systematically modify the experimental conditions and monitor the influence on the pumping, decay and dephasing rates. Firstly, we repeated the measurements presented in Fig. 1 for different excitation power densities. The normalized spectra recorded close to zero detuning ($\delta\sim0$ $\mu$eV) are plotted in Fig. 2a with excitation power densities ranging from 6 - 400 W/cm$^{2}$. As in Fig. 1, the open blue circles correspond to the experimental data whilst the solid black lines are fits. %Unfortunately, the fitting algorithm has some problems fitting the curve for the lowest excitation power due to the bad signal to noise ratio, but we can at least still notice a qualitative agreement with the experimental data.
We observe a splitting at low excitation powers into two polariton peaks. These peaks clearly merge and are not well resolved at high excitation powers. 

\begin{figure}[t!]
\includegraphics[width=86.3mm]{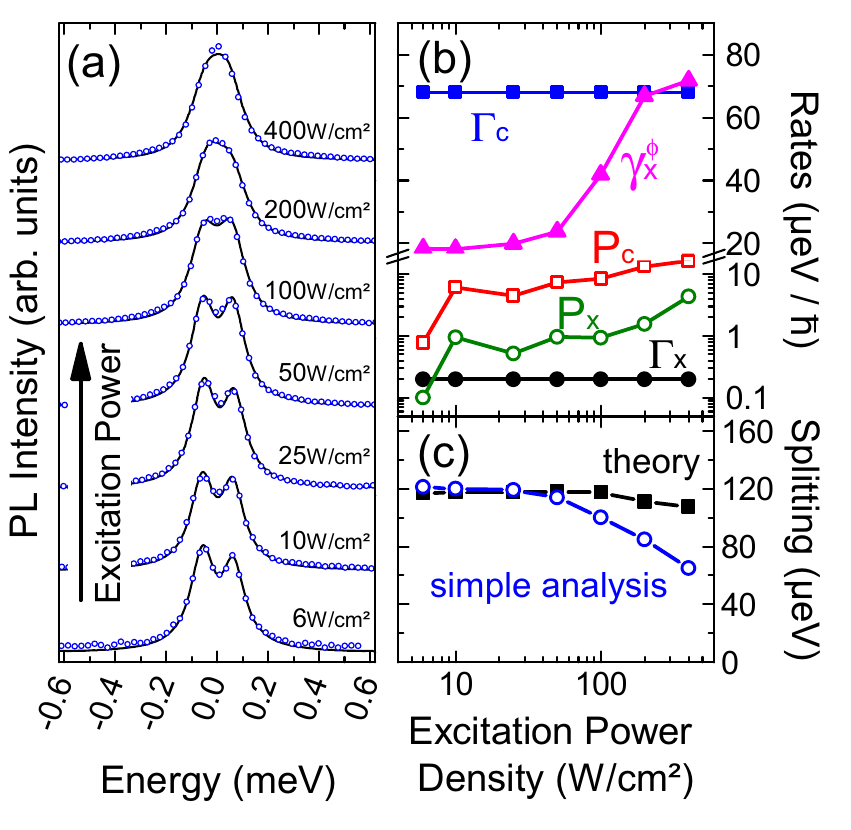}
\caption{\label{figure3} (color online) \textbf{(a)} Spectra of the exciton polaritons at resonance as a function of excitation power. The circles correspond to experimental data, while the solid lines are fits to the theory. In \textbf{(b)} the fitting parameters are displayed as they change with increasing excitation power, and the diagram in \textbf{(c)} shows the theoretically expected Rabi splitting at resonance (filled black squares) and the observed position of the peaks in the PL spectra (open blue circles).}
\end{figure}

The parameters corresponding to the fits of these data sets are summarized in Fig. 2b. We reiterate that the set of parameters for each excitation power density is obtained by \emph{globally} fitting the entire set of $\delta$-dependent data, as in Fig. 1. The green (open circles) and red (open squares) curves show the QD ($P_{x}$) and cavity mode ($P_{c}$) pumping rates, respectively. Both exhibit a clear linear and monotonous increase as expected since $P_{x}$ and $P_{c}$ should scale with the excitation power density. Similarly, the black (filled circles) and the blue curves (filled squares) show the exciton decay ($\Gamma_{x}$)and cavity loss rates ($\Gamma_{c}$), respectively. These were kept constant since they depend, respectively, on the dipole moment of the dot and the radiation losses of the cavity, but not the excitation power. Furthermore, $\hbar\Gamma_{x}=0.2$ $\mu$eV corresponds to a radiative lifetime of $\tau=3$ ns, in good agreement with time resolved measurements performed on similar QDs\cite{Kaniber07} and $\hbar\Gamma_{c}=68$ $\mu$eV corresponds to a cavity Q of $\sim17600$, close to the measured value of $Q>11500$ for this mode, bearing in mind the limited spectral resolution of the setup. Thus, the values of $\hbar P_{x}$, $\hbar P_{c}$, $\hbar\Gamma_{x}$ and $\hbar\Gamma_{c}$ extracted from the fits have physically meaningful values and exhibit precisely the expected dependence on excitation power. These observations strongly support the validity of our model.  

The only additional parameter obtained from fitting our data is the dephasing rate $\hbar\gamma_{x}^\phi$, as shown in Fig. 2b by the magenta filled triangles. $\hbar\gamma_{x}^\phi$ does not vary for low excitation powers $<$50 W/cm$^{2}$, remaining close to $\hbar\gamma_{x}^\phi\sim20\mu$eV, but increases rapidly for higher excitation powers. This observation is clear evidence for the presence of excitation induced dephasing, as was recently reported for excitons in self-assembled QDs\cite{Berthelot06,Favero07}. Here, spectral wandering induced by fluctuations in the charge status of the QD environment was shown to result in a linear temperature dependence and a characteristic power dependence, very similar to the $\hbar\gamma_{x}^\phi$ data presented in Fig. 2b. Physically, the power induced broadening arises from a transition between a regime where motional narrowing of the fluctuating environment takes place (low power), to one where the dephasing is dominated by the fluctuating QD environment\cite{Berthelot06}. Given the proximity of the investigated QD to the etched surfaces of the photonic crystal it is, perhaps, hardly surprising that excitation induced dephasing plays a significant role.

\begin{figure}[t!]
\includegraphics[width=86.3mm]{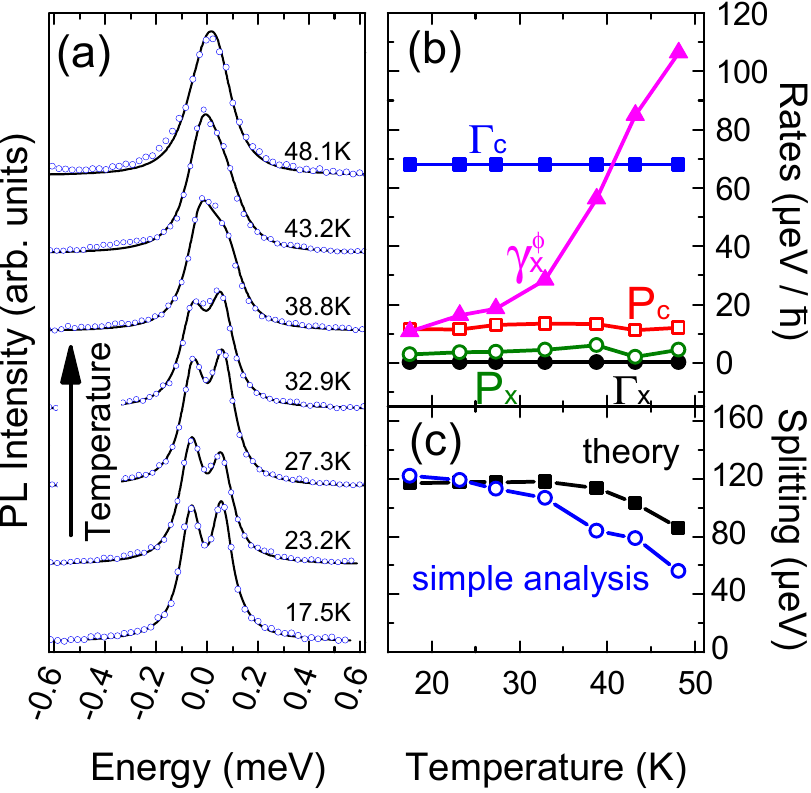}
\caption{\label{figure4}  (color online) \textbf{(a)} Spectra of the exciton polaritons at resonance as a function of temperature. The circles correspond to experimental data, while the solid lines are fits to the theory. In \textbf{(b)} the fitting parameters are displayed as they change with increasing temperature, and the diagram in \textbf{(c)} shows the theoretically expected Rabi splitting at resonance (filled black squares) and the observed position of the peaks in the PL spectra (open blue circles).}
\end{figure}

In Fig. 2c we plot the effective vacuum Rabi splitting ($2\Omega$ - Eqn. 3) as a function of excitation power density. The open blue circles correspond to the result of fitting two Lorentzian peaks to the experimental spectra as was done in Refs. \cite{Yoshie04,Reithmaier04}, whilst the filled black squares are the values obtained from the theoretical fits. The simple analysis would indicate that the vacuum Rabi frequency reduces at higher excitation levels, whereas the full theory shows that $2\Omega$ is largely unaffected by excitation. This discrepancy between a simple and full analysis was already highlighted in Refs \cite{Laussy08a,Laussy08b}, SC often appears \textquotedblleft in disguise\textquotedblright of a single peak. From the high power spectrum (400 W/cm$^{2}$) in Fig. 2a, the simple treatment would indicate that the system is not longer in the SC regime whilst our theoretical analysis indicates that the system is still in the SC regime, albeit with significant broadening due to pure dephasing. 

%TEXT FOR FIGURE 4
Besides excitation induced dephasing, another major source of decoherence in QDs is due to coupling to the lattice. Therefore, we investigated the spectrum of the QD-polaritons as a function of temperature. Examples of the recorded spectra close to $\delta\sim0$ $\mu$eV are plotted in Fig. 3a for temperatures ranging from 17.5 - 48.1 K.  The excitation power density used for these measurements was 10 W/cm$^{2}$ in the low power regime of Fig. 2. Clearly we resolve the two polariton peaks for $T<30$ K but they broaden rapidly between $30$ - $40$ K and merge into a single unresolved feature at higher temperatures. Again, we used our model to fit the entire $\delta-$dependent data at each temperature and the resulting parameters are summarized in Fig. 3b. In contrast to the power dependent measurements discussed above (Fig. 2) the temperature is not expected to influence either the decay ($\Gamma_x$,$\Gamma_c$) \emph{or} pumping ($P_x,P_c$) rates of QD and cavity mode.  This expectation is confirmed by the results of our fitting, where $\Gamma_{x/c}$ were kept constant and $P_{x/c}$ remain constant as $T$ varies.  The only fit parameter that varies strongly with the temperature is the pure dephasing rate $\hbar\gamma_{x}^\phi$ (filled magenta triangles - Fig. 3b) that increases linearly for temperatures below $30$ K with a temperature coefficient $\alpha_0\sim1.1$ $\mu$eV/K, and more rapidly for higher temperatures. The linear temperature dependence of the dephasing rate is strong evidence for decoherence mediated by coupling to acoustic phonons\cite{Besombes01,Favero03,Favero07} where reported temperature coefficients of the zero-phonon exciton transition lie in the range $\alpha_0=0.04-4$ $\mu$eV/K\cite{Urbaszek04,Borri05}. \\
Finally in Fig. 3c, we plot the experimentally polariton peak splitting obtained using the simple model (open blue circles) and from our theory (filled black squares). The effective vacuum Rabi splitting is insensitive to the temperature for $T<30$ K and reduces rapidly at higher temperature. As discussed earlier, the effective vacuum Rabi splitting differs strongly from the simple analysis of the peak splitting in all cases except conditions of weak pumping and zero temperature, underscoring the need for a full theoretial description of the emission spectrum in order to obtain an understanding of QD microcavity polaritons.\\

%Conclusion
In conclusion, we reported the first systematic invesigations of the power and temperature dependence of exciton-polaritons of a single semiconductor QD embedded within a microcavity.  By fitting the experimental data with an extended theoretical model that includes pure dephasing, we extracted the dephasing rate of the exciton polaritons and its dependence on excitation power and lattice temperature.  The results obtained illustrated how excitation induced dephasing and coupling to acoustic phonons results in dephasing of the QD-polaritons.  Most surprisingly, strong coupling is maintained even in the situation where only a single peak is observed in the emission spectrum as predicted theoretically by Ref. \cite{Laussy08a}.    

%Acknowledgements
We acknowledge financial support of the DFG via the SFB 631, Teilprojekt B3 and the German Excellence Initiative via NIM. JMVB acknowledges the support of the Alexander von Humboldt Foundation.\\

\bibliography{Papers}% Produces the bibliography via BibTeX.

\end{document}